\documentclass[sigconf]{acmart}

\AtBeginDocument{%
  }

\setcopyright{acmlicensed}
\copyrightyear{2018}
\acmYear{2018}
\acmDOI{XXXXXXX.XXXXXXX}

\acmConference[Conference acronym 'XX]{Make sure to enter the correct
  conference title from your rights confirmation emai}{June 03--05,
  2024}{NY}
\acmISBN{978-1-4503-XXXX-X/18/06}

\usepackage{xspace}

\usepackage{amsmath}
\usepackage{algorithm}
\usepackage{algpseudocode}
\usepackage{amsfonts}

\newcommand{\datasetFont}{\text}
\newcommand{\ours}{\datasetFont{\textbf{ShadeRouter}}\xspace}
\begin{document}

\title{Shaded Route Planning Using Active Segmentation and Identification of Satellite Images}

\author{Longchao Da}
\affiliation{%
  \institution{Arizona State University\\
  {longchao@asu.edu}}
  \city{Tempe, AZ}
  \country{USA}
}

\author{Rohan Chhibba}
\affiliation{%
  \institution{Arizona State University\\
  {rchhibba@asu.edu}}
  \city{Tempe, AZ}
  \country{USA}
}

\author{Rushabh Jaiswal}
\affiliation{%
  \institution{Arizona State University\\
  {rjaisw15@asu.edu}}
  \city{Tempe, AZ}
  \country{USA}
}

\author{Ariane Middel}
\affiliation{%
  \institution{Arizona State University\\
  {ariane.middel@asu.edu}}
  \city{Tempe, AZ}
  \country{USA}
}

\author{Hua Wei$^*$}
\affiliation{%
  \institution{Arizona State University\\
  {hua.wei@asu.edu}}
  \city{Tempe, AZ}
  \country{USA}
}

\renewcommand{\shortauthors}{Trovato et al.}

\begin{abstract}
Heatwaves pose significant health risks, particularly due to prolonged exposure to high summer temperatures. Vulnerable groups, especially pedestrians and cyclists on sun-exposed sidewalks, motivate the development of a route planning method that incorporates somatosensory temperature effects through shade ratio consideration. This paper is the first to introduce a pipeline that utilizes segmentation foundation models to extract shaded areas from high-resolution satellite images. These areas are then integrated into a multi-layered road map, enabling users to customize routes based on a balance between distance and shade exposure, thereby enhancing comfort and health during outdoor activities. Specifically, we construct a graph-based representation of the road map, where links indicate connectivity and are updated with shade ratio data for dynamic route planning. This system is already implemented online, with a video demonstration, and will be specifically adapted to assist travelers during the 2024 Olympic Games in Paris.
\end{abstract}


\maketitle

\section{Introduction}
The impact of global warming is increasingly evident worldwide. In some regions, the number of deaths related to high temperatures is alarming. A study by \cite{monashstudy} reports that, from 2000 to 2019, an average of 178,700 deaths annually were attributed to high temperatures, with the number continuing to rise yearly. The escalating severity and frequency of heatwaves, exacerbated by climate change, represent a significant public health threat. A more accessible, comfortable, and safe route planning method aligns as a key priority with both the European Union's Green Deal~\cite{fetting2020european} and the United Nations Sustainable Development Goals~\cite{bexell2017responsibility}.

There are some literature that conducted preliminary research on the possibility of planning under shaded areas. Such as~\cite{dostal2009quantifying} proposes to use BOTworld to find and visualize optimal thermal comfort paths in squares, parks, and small neighborhoods, such an agent-based model acts with ENVImet microclimate~\cite{liu2021heat} simulations. ~\cite{vasilikou2020outdoor} investigated \textit{thermal walks} in two European pedestrian routes using questionnaires and field measurements to improve models of dynamic thermal comfort perception. In the study of ~\cite{vartholomaios2023follow}, researchers propose a novel way to detect optimal pedestrian shaded paths using UMEP~\cite{lindberg2015umep} in QGIS~\cite{qgis}, however, this solution relies on accurate terrain maps, and real-time building heights are not well-considered. Besides,~\cite{lidarPath}, proposes a method to conduct shaded navigation using a cross-source of OpenStreetMap~\cite{bennett2010openstreetmap} and LiDAR point cloud data that resolves both treetop canopy and bare ground elevation problems, however, this relies on well-collected LiDAR data~\cite{liu2007effect} and is hard to apply directly to arbitrary cities.

In this paper, we formally present \ours, a novel shaded route planning method, and provide a demo to showcase its real-time planning ability. 
This paper contributes in the following aspects: 
\textbf{First}, this paper introduces a complete pipeline leveraging the advanced foundation model to extract shaded information from satellite images, which can be directly applied to any city with available satellite information. \textbf{Second}, the paper proposes a shaded ratio calculation algorithm, which takes the satellite image and OpenStreetMap as input, and derives the percentage of shade-covered length to the whole route length. Such shaded ratios will be adopted to create an information graph that will be jointly considered with a distance graph when providing routing plans. \textbf{Third}, we provide an online demo for route planning employing a variant of the Dijkstra algorithm and release the source-code, dataset, and corresponding derived shaded ratios for other researchers' convenience.


\begin{figure*}[h!]
    \centering
    \includegraphics[width=0.90\textwidth]{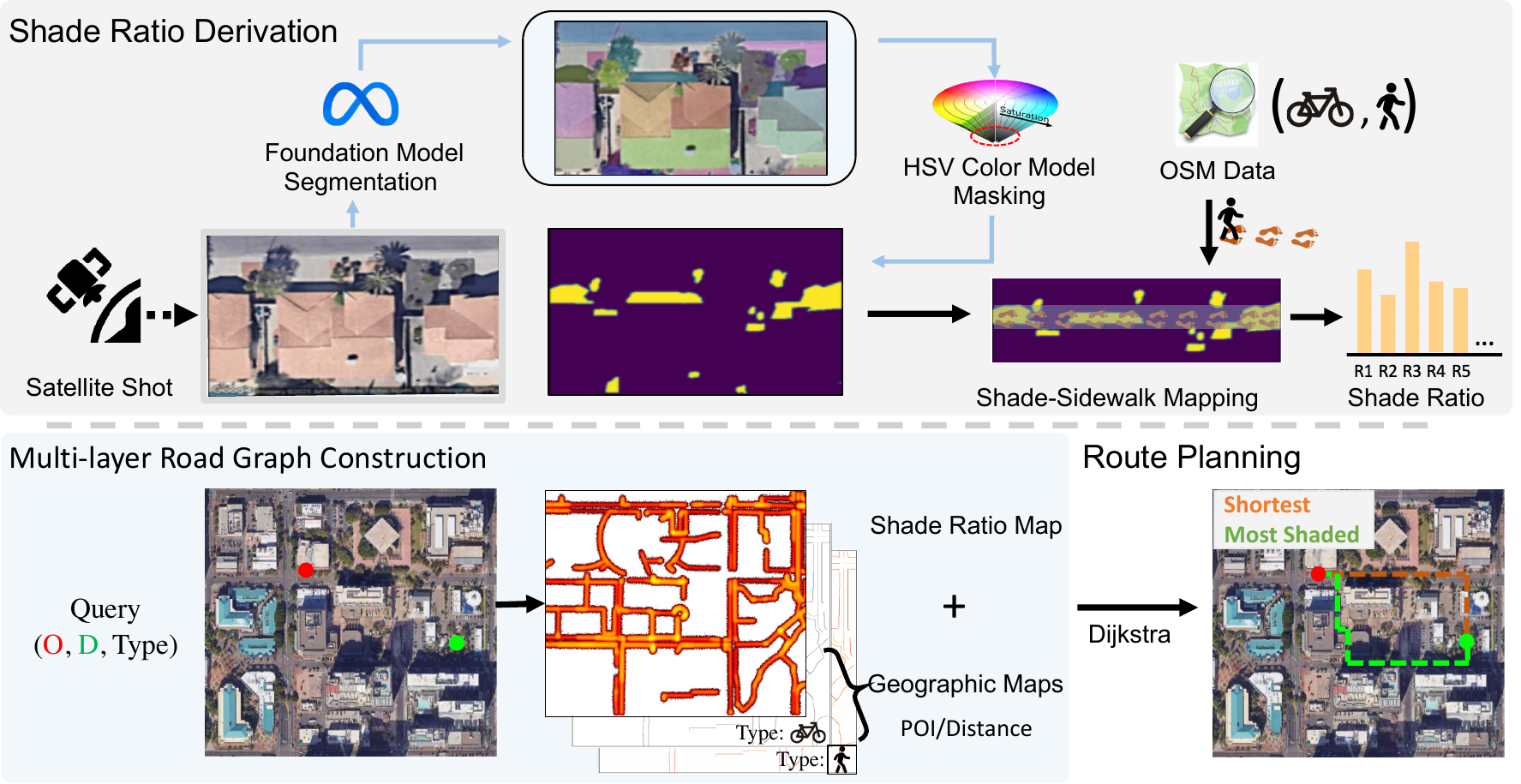}
    \caption{The overview of the proposed pipeline for our shaded route planning method. The upper part shows the shaded ratio derivation process that takes the satellite image and OSM data as input, and calculates the shaded ratio for specific valid walkable or bikeable lanes. And the lower part shows the multi-layer road graph construction and route planning process, this reveals how \ours provides user preference shaded route planning from derived shadow information.}
    \label{fig:overview}
\end{figure*}



\section{Approach}
In this section, we will introduce the proposed planning method \ours. The detailed pipeline includes three components: \textit{Shade Ratio Derivation}, \textit{Multi-layer Road Graph Construction}, and \textit{Route Planning Interface}. Previous studies have attempted to utilize LiDAR data~\cite{wang2020estimation, behrendt2005temperature, leblanc1998temperature} to simulate shade and enhance route planning~\cite{buo2023high,vartholomaios2023follow}. However, these methods often struggle with localization issues and data sparsity for places without LiDAR data, limiting their applicability to certain regions. Our approach leverages widely available satellite imagery to derive shade information globally. This data source provides detailed environmental features, such as vegetation and buildings, ensuring our shade-based navigation planning is both versatile and reliable.

\subsection{\textbf{Shade Ratio Derivation}}
This step utilizes satellite imagery to derive information about shaded areas on maps, as illustrated in the first module of Fig.~\ref{fig:overview}. The process begins with the segmentation of raw satellite images using the foundation model, SegmentAnything~\cite{kirillov2023segment, ke2024segment, jing2023segment}, which employs contrastive learning to identify and delineate each object within the image. 
Once segmented, the image components are analyzed for color hue and brightness to determine shaded areas. This is achieved through chromatics analysis, which identifies darker areas as shaded. In our empirical study, the best threshold value for selecting RGB masks is 75 (mask keeps iff RGB(mask) >= 75). Subsequently, these shaded areas are aligned with road data from OpenStreetMap~\cite{bennett2010openstreetmap}. By overlaying the shade map onto valid path coordinates, as shown in Fig.~\ref{fig:overview}, we can determine the shade ratio, i.e., the percentage of each road covered by shade, facilitating the identification of walkable shaded lanes. 

\begin{figure}
    \centering
    \includegraphics[width=0.49\textwidth]{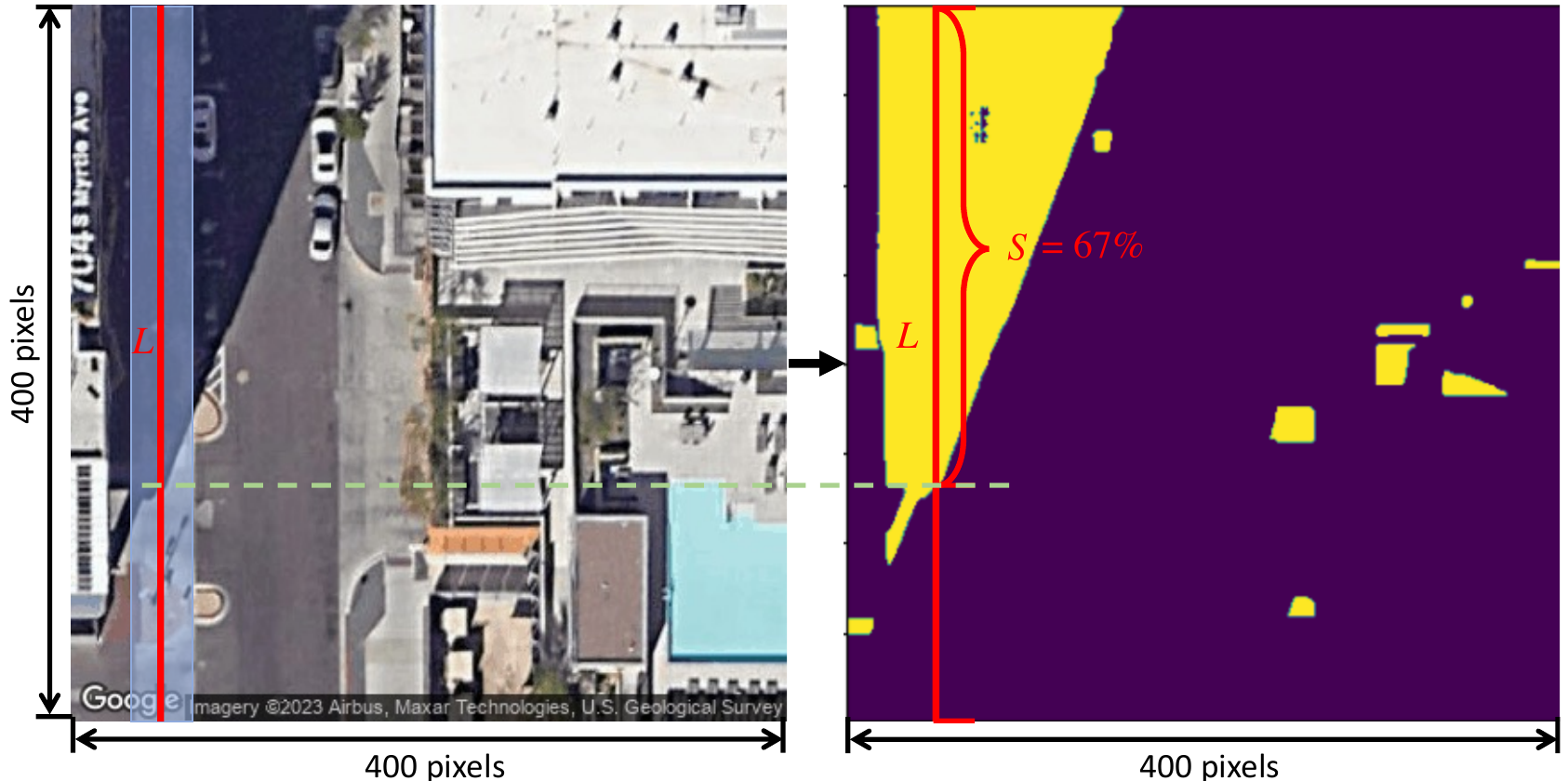}
    \caption{The shaded ratio calculation, the yellow blocks show the shaded areas, and the red line shows valid routes from OSM data.}
    \label{fig:ratio}
\end{figure}

The Fig.~\ref{fig:ratio} shows another example of shaded ratio calculation in a walkable network, in this yellow color masked satellite image (such mask implies the shaded area), the red line is a valid pedestrian route extracted from the OSM files~\cite{lu2023virtual}, as shown in the image, the total pixel $L= 400$, by image processing, we calculate the overlap between valid pedestrian route and the actual shaded area is $S=67\% \times L$. in our setting, each image is downloaded with a zoomed-in level as 20, result in the range of 49.84m x 49.84m. Given this information, we could calculate the shaded ratio for the above example in Fig.~\ref{fig:ratio} which is 49.84 $\times$ 67\% $\approx$ 33.39m. 

Another challenge in the shade ratio derivation is, the dataset of a place contains multiple images, and the same route may appear multiple times in different images.  In order to deal with this, we take the road name as the key (K), assume we already have partially detected shaded length as $L_{Shaded}$, accumulated length as $L_{Acc}$, and now we conduct query Q($img_i$, K) in each un-visited dataset image, if Q($img_i$, K) = True, indicating a positive query, the $L_{Shaded} = L_{Shaded} + L_{Shaded}^{K}$, and $L_{Acc} = L_{Acc} + L_{Acc} ^ {K}$. The final shaded ratio for route $K$ is $r(K) = \frac{L_{Shaded}}{L_{Acc}}$. 
If we traverse all interested routes, we could derive a pre-processed shaded ratio map.

\subsection{\textbf{Multi-layer Road Graph Construction}}
Following the derivation of shade ratio data, we employ this data in the construction of a \textit{Shaded Ratio Map}, as part of our Multi-layer Road Graph Construction shown in Fig.~\ref{fig:overview}. This construction comprises two primary layers: the upper layer represents the shade ratios derived from our earlier process, and the lower layer represents geographic maps, denoted as $\mathbf{\mathcal{G}}_{Type}$ (where ${Type}$ is either \textit{walk} or \textit{bike}), sourced from OpenStreetMap (OSM) data. It is important to note that the connectivity in $\mathbf{\mathcal{G}}_{walk}$ and $\mathbf{\mathcal{G}}_{bike}$ may differ. The shaded ratio layer acts as a universal set, $\mathbf{\mathcal{G}_{\mathcal{U}}}$, encompassing all possible links. While mapping the shaded ratio map to the geographical maps, if the connectivity does not exist, then it will be automatically removed from consideration because it will be identified as \textit{in-accessible}. After mapping all of the route vertices and links, well-defined data sources are available for various route planning methods in the next steps. 


\begin{figure*}[h!]
    \centering
    \vspace{-3mm}   
    \includegraphics[width=1.0\textwidth]{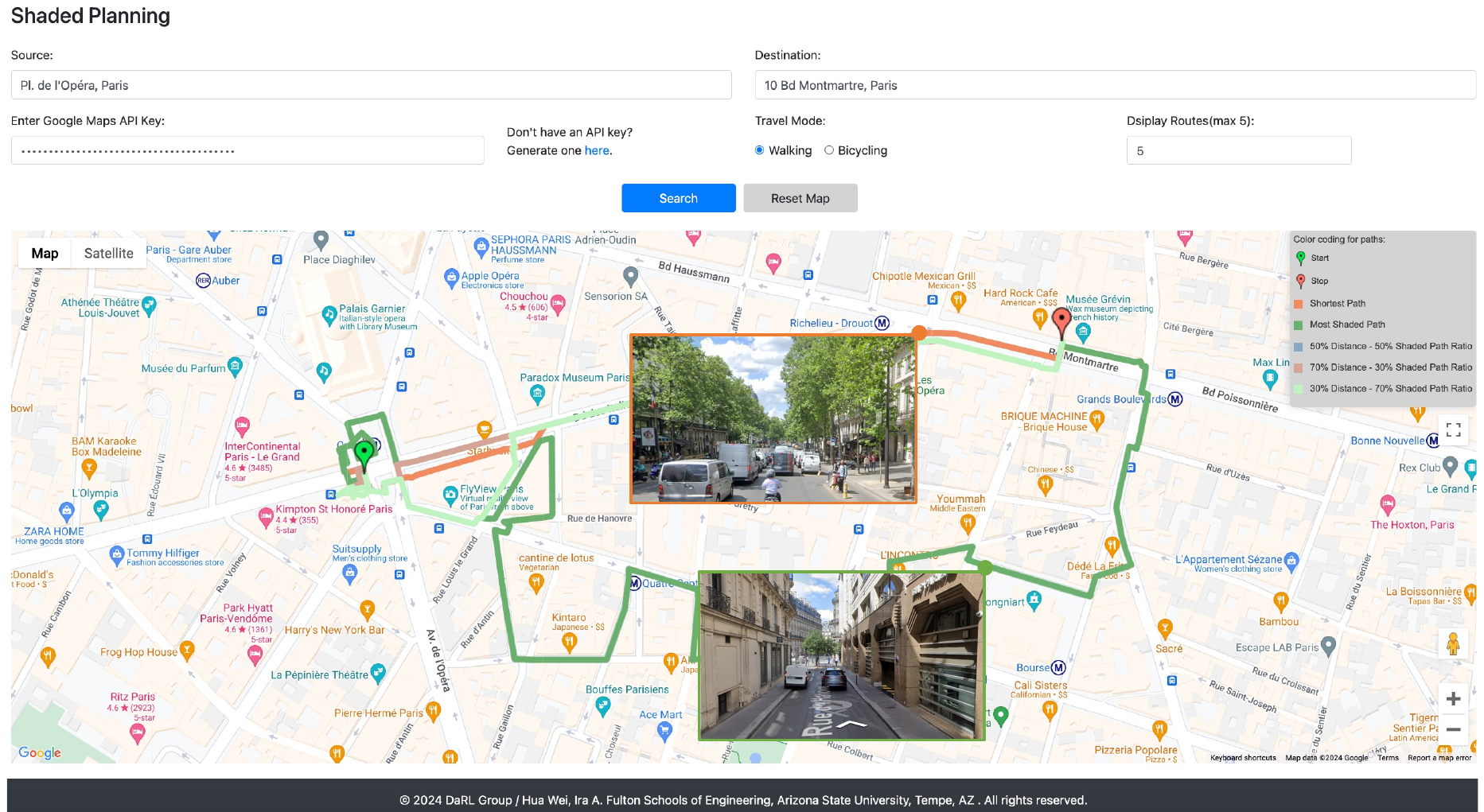}
    \caption{The comparison between two itineraries by \ours in Paris. The shortest route in orange color is more exposed to heatwaves compared to the most shaded route, as shown in the snapshot of street views.}
    \label{fig:comparison}
\end{figure*}

\subsection{\textbf{Route Planning and User Interface}}
Equipped with the road graph and shaded ratio graph, our system can now offer customizable routing recommendations through a modified version of the Dijkstra~\cite{noto2000method} algorithm. Users can input a query in the format $\textit{Query}(O, D, Type)$, where $\textit{O}$ and $\textit{D}$ represent the coordinates of the starting points and destinations, respectively, denoted as ($\textit{x, y}$) and ($\textit{x}', \textit{y}'$). The $\textit{Type}$ parameter specifies the mode of transportation, either \textit{walk} or \textit{bike}. Depending on the selected type, different graphs are utilized: $\textit{T}_{walk}$ uses $\{\mathbf{\mathcal{G}}_{walk}, \mathbf{\mathcal{G}_{\mathcal{U}}}\}$, and $\textit{T}_{bike}$ employs $\{\mathbf{\mathcal{G}}_{bike}, \mathbf{\mathcal{G}_{\mathcal{U}}}\}$. This system allows for route planning that is tailored to the user’s specific preferences and needs. Let $\alpha$ be the preference for a more shaded route $\alpha \in [0, 1]$, and $1-\alpha$ would be the preference weight for the distance. Thus, for any edge with connectivity, the value of the edge will be updated by the equation: 
\begin{equation}
    w_{joint} (u, v) = \alpha V_{shade} + (1 - \alpha) V_{distance}
\end{equation}
where the $V_{shade}$ can be represented as $r(u, v)$, given the two point of interests $u$ and $v$, and $r(\cdot)$ as the shade ratio calculation function. 

\begin{algorithm}
\caption{\ours Planning}
\begin{algorithmic}[1]
\Require Graph $G = (V, E)$, origin vertex $v_o$, destination vertex $v_d$, shaded ratios $r(u, v)$ for each edge $(u, v)$, preference parameter $\alpha \in [0, 1]$
\State Initialize $d[v] \gets \infty$ for all $v \in V$
\State $d[v_o] \gets 0$
\State Initialize priority queue $Q$
\ForAll{vertex $v \in V$}
    \State Insert $v$ into $Q$ with priority $d[v]$
\EndFor
\State Initialize $prev[v] \gets \text{null}$ for all $v \in V$
\While{$Q$ is not empty}
    \State $u \gets $ Extract vertex with minimum distance from $Q$
    \ForAll{neighbor $v$ of $u$}
        \State $w_{\text{joint}}(u, v) \gets (1 - \alpha) \cdot w(u, v) + \alpha \cdot r(u, v)$
        \If{$d[u] + w_{\text{joint}}(u, v) < d[v]$}
            \State $d[v] \gets d[u] + w_{\text{joint}}(u, v)$
            \State $prev[v] \gets u$
            \State Decrease priority of $v$ in $Q$ to $d[v]$
        \EndIf
    \EndFor
\EndWhile
\State Construct path $P \gets []$
\State $u \gets v_d$
\While{$prev[u] \neq \text{null}$}
    \State \text{insert } $u$ \text{ at the beginning of } $P$
    \State $u \gets prev[u]$
\EndWhile
\State \text{insert } $v_o$ \text{ at the beginning of } $P$
\State \Return $P$
\end{algorithmic}
\end{algorithm}

In the \textbf{algorithm 1}, the $V$ is the set of interested points from the OSM-filtered bikeable or walkable graph network, and $E$ is the set of accessible lanes (edges), and after the calculation, the returned $P$ contains the $Path$ of the suggested plan, which is a list of POIs and can be used for planning and navigation. Alternatively, one can easily modify the above algorithm and output the top $k$ suggested paths based on their different preference $\alpha$ reflecting on a balance of shaded areas and distance.

\section{Dataset Construction}
While conducting route planning tasks, in this paper, we get rid of the localized specially collected LiDAR data, instead, we propose to adopt a more accessible resource - satellite image as base data, and we query the Google Map API~\footnote{\url{https://developers.google.com/maps/apis-by-platform}} for the satellite images of a specific city in the demo. We use the 20x zoomed pile image data for processing, the resolution at the equator is 0.1246 meters per pixel (there are 400 pixels in total, so the range in equator is 400 $\times$ 0.1246 = 49.84 m). To a general latitude, the resolution $Res$ is: 
\begin{equation}
    Res=49.84 \times Cosin(lat)
\end{equation}
where the $lat$ is the latitude in degrees.

\begin{table}[h!]
  \caption{The dataset statistics and file information.}
  \label{tab:params}
  \begin{tabular}{ccccc}
    \toprule
    Dataset &Images & Size & Latitude-Longitude Range \\  
    \midrule
    Tempe & 38,796 & 3.67 GB & (33.43, 33.32, -111.97, -111.89) \\
    Paris & 26,052 & 2.94 GB & (48.88, 48.83,  2.30, 2.39) \\
    Byeng & 72 & 6.6 MB & (33.425, 33.422, -111.941, -111.936) \\
  \bottomrule
\end{tabular}
\end{table}

The data is released at the link of the footnote ~\footnote{\url{https://www.dropbox.com/scl/fo/rvq0nazh7sd4qlx013l16/ANzGN1kQIvj88Mrpc7m6-w8?rlkey=w3xml0kv75hbt80gz8km75phd&dl=0}}. In these datasets, we covered two complete cities \texttt{Paris, France} and \texttt{Tempe - AZ, USA}. We also include a case study dataset \texttt{Byeng} which is the location of the Brickyard Engineering Building, in Tempe, AZ, No.699.

\section{Demo Presentation}
In this section, we introduce the demo program, the recorded demo video~\footnote{\url{https://drive.google.com/drive/folders/1BvR0i0eImDb86HXiGS4hGRoFKa71DAnJ?usp=sharing}} and provide an introduction to the \ours. 

In the implemented program, the user interface incorporates 4 text boxes, two buttons, and one travel mode selector, with a map viewing area at the bottom. Corresponding to the user input, the user would be required to input the source point and destination point, the extra two boxes are for the Google map API and the number of suggested users intend to receive. Once two valid places are searched, the system will show up top $\textit{k}$ marked routes and pop up a legend explaining each color's meaning. The two images shown in Fig.~\ref{fig:comparison}. are from two results, the shortest path, and the most shaded path, we could notice that the orange line gives the fastest itinerary with high exposure to sunshine, and the green one prefers alley that is more shaded and thermally comfortable. This enables users to find the most preferred plans that improve the travel experience. 
Despite the demo, for more interaction, please visit the deployed website~\footnote{\url{https://longchaoda.github.io/ShadedPlanning.github.io/}} and it is also welcome to check our released code site ~\footnote{\url{https://github.com/LongchaoDa/Shaded_Planning.git}}.

\section{Conclusion}
In this work, we propose a prototype method of \ours, which leverages the satellite image to mine the shaded ratio. By doing so, it provides the route planning suggestions for pedestrians and cyclists considering the shades on the road. It is feasible to extend the demo to a worldwide map using the formally designed pipeline based on the publicly accessible satellite image data.

This demonstration can be further improved by temporal information such as the time or seasonal shade situation. We hope this project will benefit the citizens' health by reducing the heatwave exposure probability and providing a more comfortable outdoor activity plan.

Future developments include adapting the current method to a more dynamic shade simulation that captures the real-time shadow changes in cities, and optimizes the planning speed to enhance the users' experience. If equipped with more sensible data, or more outdoor choices, the algorithm could even include wind strength and extreme weather conditions in the planning, and provide e-scooters and shared bikes as alternatives, and these will definitely provide richer activity choices for city-travelers.

\bibliographystyle{ACM-Reference-Format}
\bibliography{sample-base}

\appendix

\end{document}